\documentclass{elsart_mod}
\usepackage[square,numbers,comma,sort&compress]{natbib}
\usepackage{amssymb,amsmath,amsfonts}
\usepackage{graphicx,here}

\DeclareMathOperator{\erf}{Erf} \DeclareMathOperator{\erfi}{Erfi}

\begin{document}

\runauthor{Ellis, Mavromatos, Mitsou and Nanopoulos}
\begin{frontmatter}
\title{Confronting Dark Energy Models with Astrophysical Data: Non-Equilibrium
vs.\ Conventional Cosmologies}
\author[Ellis]{John R.\ Ellis},
\author[Mavromatos]{Nikolaos E.\ Mavromatos},
\author[Mitsou]{Vasiliki A.\ Mitsou},
\author[Nanopoulos1,Nanopoulos2,Nanopoulos3]{Dimitri V.\ Nanopoulos}

\address[Ellis]{TH Division, Physics Department, CERN, CH-1211 Geneva
23, Switzerland}
\address[Mavromatos]{Theoretical Physics, Physics Department,
King's College London, \\
Strand WC2R 2LS, UK}
\address[Mitsou]{Instituto de F\'{i}sica Corpuscular (IFIC), CSIC --
Universitat de Val\`{e}ncia, \\
Edificio Institutos de Paterna, P.O. Box 22085, E-46071
Valencia, Spain}
\address[Nanopoulos1]{George P.\ and Cynthia W.\ Mitchell Institute for
Fundamental Physics, \\
Texas A\&M University, College Station, TX~77843, USA}
\address[Nanopoulos2]{Astroparticle Physics Group, Houston Advanced Research
Center (HARC), Mitchell Campus, Woodlands, TX~77381, USA}
\address[Nanopoulos3]{Academy of Athens, Division of Natural Sciences,
28~Panepistimiou Avenue, Athens GR-10679, Greece}

\begin{abstract}
We discuss fits of cosmological dark energy models to the available data on
high-redshift supernovae. We consider a conventional model with Cold Dark
Matter and a cosmological constant ($\Lambda$CDM), a model invoking
super-horizon perturbations (SHCDM) and models based on Liouville strings in
which dark energy is provided by a rolling dilaton field (Q-cosmology). We show
that a complete treatment of Q-cosmology requires a careful discussion of
non-equilibrium situations (off-shell effects). The two main high-redshift
supernova data sets give compatible constraints on $\Lambda$CDM and the other
models. We recover the well-known result that $\Lambda$CDM fits very well the
combined supernova data sets, as does the super-horizon model. We discuss the
model-dependent off-shell corrections to the Q-cosmology model that are
relevant to the supernova data, and show that this model fits the data equally
well. This analysis could be extended to other aspects of cosmological
phenomenology, in particular to the CMB and Baryon Acoustic Oscillations, which
have so far been treated using on-shell models.
\end{abstract}


\end{frontmatter}

\section{Introduction}

There is a plethora of astrophysical evidence today, from supernovae
measurements, the spectrum of fluctuations in the cosmic microwave background
\cite{snIa,wmap:s,steinhardt,deceldata}, baryon oscillations and other
cosmological data, indicating that the expansion of the Universe is currently
accelerating. The energy budget of the Universe seems to be dominated at the
present epoch by a mysterious dark energy component, but the precise nature of
this energy is still unknown. Many theoretical models provide possible
explanations for the dark energy, ranging from a cosmological constant
\cite{concordance} to super-horizon perturbations \cite{riotto} and
time-varying quintessence scenarios \cite{steinhardt}, in which the dark energy
is due to a smoothly varying (scalar) field which dominates cosmology in the
present era, such as a time-dependent dilaton field \cite{aben}.

The current astrophysical data are capable of placing severe constraints on the
nature of the dark energy, whose equation of state may be determined by means
of an appropriate global fit. Most of the analyses so far are based on
effective four-dimensional Robertson-Walker Universes, which satisfy on-shell
dynamical equations of motion of the Einstein-Friedman form. Even in modern
approaches to brane cosmology, which are described by equations that deviate
during early eras of the Universe from the standard Friedman equation (which is
linear in the energy density), the underlying dynamics is assumed to be of
classical equilibrium (on-shell) nature, in the sense that it satisfies a set
of equations of motion derived from the appropriate minimization of an
effective space-time Lagrangian.

However, cosmology may not be an entirely classical equilibrium situation
\cite{aben}.  The initial Big Bang or other catastrophic cosmic event, which
led to the initial rapid expansion of the Universe, may have caused a
significant departure from classical equilibrium dynamics in the early
Universe, whose signatures may still be present at later epochs including the
present era. Three of us (JE, NEM, DVN) have indeed proposed one specific model
for the cosmological dark energy which is of this type, being associated with a
rolling dilaton field that is a remnant of this non-equilibrium phase,
described by a generic non-critical string theory
\cite{emnw,diamandis,diamandis2}. We call this scenario `Q-cosmology'.

The central purpose of this paper is to confront the present cosmological data
on high-redshift supernovae with non-equilibrium cosmologies \cite{aben, emnw,
diamandis, diamandis2}, in which the dark energy relaxes at different rates,
and compare the results with the predictions of the conventional $\Lambda$CDM
model \cite{concordance}. We also comment on a model with super-horizon
perturbations superposed on an underlying Einstein-Friedman-Robertson-Walker
Universe\footnote{Our approach to this model is strictly phenomenological: we
do address any of its deeper theoretical issues.} \cite{riotto}.

As we explain in more detail below, care must be taken in interpreting the
Q-cosmology scenario proposed in Refs.~\cite{emnw,diamandis,diamandis2}. Since
such a non-equilibrium, non-classical theory \emph{is not described by the
equations of motion derived by extremizing an effective space-time Lagrangian},
one must use a more general formalism to make predictions that can be
confronted with the current data. The approach we favour is formulated in the
context of string/brane theory \cite{gsw,polchinski}, the best candidate theory
of quantum gravity to date. Our approach is based on non-critical (Liouville)
strings \cite{aben,ddk,emn}, which offer a mathematically consistent way of
incorporating time-dependent backgrounds in string theory.

The basic idea behind such non-critical Liouville strings is the following.
Usually, in string perturbation theory, the target space dynamics is obtained
from a stringy $\sigma$-model~\cite{gsw} that describes the propagation of
strings in classical target-space background fields, including the space-time
metric itself.  Consistency of the theory requires conformal invariance on the
world sheet, in which case the target-space physics is independent of the scale
characterising the underlying two-dimensional dynamics. These conformal
invariance conditions lead to a set of target-space equations for the various
background fields, which correspond to the Einstein/matter equations derived
from an appropriate low-energy effective action that is invariant under general
coordinate transformations. Unfortunately, one cannot incorporate in this way
time-dependent cosmological backgrounds in string theory, since, to low orders
in a perturbative expansion in the Regge slope $\alpha '$, the conformal
invariance condition for the metric field would require a Ricci-flat
target-space manifold, whereas a cosmological background necessarily has a
non-vanishing Ricci tensor.

To remedy this defect, and thus be able to describe a time-dependent
cosmological background in string theory, the authors of Ref.~\cite{aben}
suggested that a non-trivial r\^ole should be played by a time-dependent
dilaton background.  This approach leads to strings living in numbers of
dimensions different from the customary critical number, and was in fact the
first physical application of non-critical strings \cite{ddk}. The approach of
Ref.~\cite{aben} was subsequently extended \cite{emn,emnw,diamandis,diamandis2}
to incorporate off-shell quantum effects and non-conformal string backgrounds
describing other non-equilibrium cosmological situations, including
catastrophic cosmic events such as the collision of two brane worlds,
\emph{etc}.

In our discussion of such models in this paper, we first review briefly their
main predictions. We then demonstrate the importance of off-shell corrections
to the naive equations of motion of non-critical Liouville string cosmology,
which cause us to modify the naive equations of motion of Q-cosmology. We then
demonstrate that the available supernova data are compatible with such
non-critical-string-based cosmologies \cite{emnw,diamandis2}.

The structure of the article is as follows. In Section~\ref{sec:pheno} we
summarise the basic equations of the first set of models we study, namely
$\Lambda$CDM and the super-horizon model. Then, in Section~\ref{sec:qcosmo} we
discuss the Liouville Q-cosmology model of
Refs.~\cite{emnw,diamandis,diamandis2}. In particular, we discuss technical
aspects of Q-cosmology, placing the emphasis on the off-shell corrections and
deriving the fitting formulae to be used in our analysis. This part may be
omitted by the reader who is more interested in the confrontation of this and
other models with the available high-redshift supernova data that we present in
Section~\ref{sec:data}. We find that the $\Lambda$CDM model fits the data very
well, as does the super-horizon model. We also show that the off-shell
corrections to the Q-cosmology enable this model to be reconciled with the
supernova data on equal footing with $\Lambda$CDM.\footnote{A more complete
comparison of this model with data on the cosmic microwave background (CMB) and
baryon acoustic oscillations (BAO) would require further formal developments
that lie beyond the scope of this paper.} Our results and the possible outlook
are summarised in Section~\ref{sec:summary}.

\section{Basic Formulae for the Phenomenological Analysis of On-Shell Dark
Energy Models}\label{sec:pheno}

In this Section we review the basic formulae needed in the phenomenological
analysis of observational constraints on the cosmological parameters derived
from supernova data and (in the case of the $\Lambda$CDM) also baryon acoustic
oscillations, in the context of on-shell models (i.e., excluding the
Q-cosmology model to be discussed later). The relevant dynamics of any such
cosmological Robertson-Walker model may be lumped into the expression for the
Hubble parameter $H(z)$ in terms of the various components of the energy
density of the Universe, represented in units of the (present-day) critical
density for a spatially-flat Universe, $\Omega_i \equiv \rho_i^0/\rho_c^0$,
where $i={\rm M}$ for matter, including dark matter, $i = \Lambda$ for constant
dark energy, i.e., a cosmological constant $\Lambda$, \emph{etc}. Later in this
work we discuss fits to the on-shell ($\Lambda$CDM and super-horizon) and
off-shell (Q-cosmology) models discussed in the current and the next Sections,
respectively.

\subsection{Standard $\varLambda$CDM Model \cite{concordance}}

In this case, we consider a spatially flat Robertson-Walker Universe containing
dust-like matter that can be described as an ideal fluid, and a cosmological
constant $\Lambda > 0$ corresponding to a \emph{constant} equation of state of
the form:
\begin{equation}
p = w_0 \rho ~,
\label{eqstate}
\end{equation}
where $p (\rho)$ is the pressure (energy density) and $w_0 = -1$. The standard
Einstein-Friedmann equations for such a Universe yield:
\begin{equation}
H(z) = H_0 \left(\Omega_{\rm M} ( 1 + z)^3 + \Omega_\Lambda ( 1 + z )^{3(1 +
w_0)}\right)^{1/2} \;, \label{hubble}
\end{equation}
where $H_0$ is the present-day value of the Hubble constant.

\subsection{Super-Horizon Perturbation Model \cite{riotto}}

In this model, the Universe is assumed to be filled with only non-relativistic
matter, and there is no Dark Energy of any sort. The metric takes the form:
\begin{equation}
ds^2 = -dt^2 + a^2(t)e^{-2\Psi({\vec x}, t)}\delta_{ij}dx^idx^j \;,
\label{kolbriotto}
\end{equation}
where $\Psi({\vec x},t)$ denotes the gravitational potential. The authors of
Ref.~\cite{riotto} considered perturbations $\Psi({\vec x},t)$ that correspond
to conformal contraction or stretching of the three-dimensional Euclidean
space, $\{x^i, i=1,2,3\}$. Splitting the potential $\Psi = \Psi_\ell + \Psi_s$
into contributions from long-wavelength (super-horizon) modes, $\Psi_\ell$, and
short-wavelength (sub-horizon) modes, $\Psi_s$, and ignoring the latter within
our Hubble volume, one obtains \cite{riotto}:
\begin{equation}
ds^2 \simeq -dt^2 + {\overline a}^2(t)\delta_{ij}dx^idx^j~,
\qquad {\overline a}(t) = a(t)e^{-\Psi_\ell(t) + \Psi_{\ell 0}}~,
\label{riottometric}
\end{equation}
with the normalization ${\overline a}_0 = a_0 = 1$.

The analysis of Ref.~\cite{riotto} used the appropriate Einstein's equations to
demonstrate that, up to trivial rescaling of spatial coordinates and absorbing
the time-independent part of $\Psi_\ell$, the time dependence of $\Psi_\ell
(t)$ may be cast in the form:
\begin{equation}
\Psi_\ell (t) \simeq a(t) \Psi_{\ell 0}.
\label{timedeppsi}
\end{equation}
One therefore finds that the expansion rate of the Universe, as observed by a
local observer who is restricted to observe within our Hubble volume, is given
by:
\begin{equation}
H(z) = {\overline a}^{-1}\frac{d{\overline a}(t)}{dt}=
\frac{{\overline H}_0}{1 - \Psi_{\ell 0}}\left(a^{-3/2} -
a^{-1/2}\Psi_{\ell 0}\right)~,
\label{modhubble}
\end{equation}
where $(1 + z)^{-1} = \overline{a}(t)$. Combining \eqref{riottometric} and
\eqref{timedeppsi}, one may extract the following relation between $a$ and $z$:
\begin{equation}
1+ z = a^{-1} e^{(a - 1)\Psi_{\ell 0}}.
\label{redshiftrelation}
\end{equation}
This model predicts acceleration of the Universe at the current era:
\begin{equation}
{q} = - 1 + \frac{3/2 - a\Psi_{\ell 0}/2}{(1 - a \Psi_{\ell 0})^2} .
\label{riottoaccel}
\end{equation}
It can easily be shown by direct comparison with conventional models that the
super-horizon perturbations mimic a dark energy model with equation of state
\begin{equation}
w_{DE}\Omega_{DE}=\frac{2}{3}\left({q} -
\frac{1}{2}\right).
\label{wDE}
\end{equation}
In order to evaluate the constraints imposed by astrophysical data on this
super-horizon model, we use the following integral of the Hubble parameter
$E(z) \equiv H(z)/H_0$ over $z$:
\begin{multline}
\int_0^{z} \frac{dz}{E(z)}
= (1 - \Psi_{\ell 0}) \int_{a(z)}^1 da~a^{-1/2}e^{(a -1)\Psi_{\ell 0}} \\
= \begin{cases}%
\sqrt{\pi}~e^{-\Psi_{\ell 0}} \frac{1 - \Psi_{\ell 0}}{\sqrt{-\Psi_{\ell 0}}}
\Bigl(\erf\bigl(\sqrt{-\Psi_{\ell 0}}\bigr) -
\erf\bigl(\sqrt{-\Psi_{\ell 0}a(z)}\bigr)\Bigr), & \mbox{if }\Psi_{\ell 0}<0 \\
2\big(1-\sqrt{a(z)}\big), & \mbox{if }\Psi_{\ell 0}=0 \\
\sqrt{\pi}~e^{-\Psi_{\ell 0}} \frac{1 - \Psi_{\ell 0}}{\sqrt{\Psi_{\ell 0}}}
\Bigl(\erfi\bigl(\sqrt{\Psi_{\ell 0}}\bigr) - \erfi\bigl(\sqrt{\Psi_{\ell
0}a(z)}\bigr)\Bigr), & \mbox{if }\Psi_{\ell 0}>0
\end{cases}\label{integral}
\end{multline}

where $\erf$ and $\erfi$ are the standard and imaginary error function,
respectively. We assume values $|\Psi_{\ell 0}| < 1$, in order for perturbation
theory to remain valid~\cite{riotto}. We reiterate that we do not address any
of the deeper theoretical issues of this model.

\subsection{A Note on Relaxing Dark Energy Models}

There are many dark energy models in which the energy density relaxes in time.
For example, relaxation at a rate proportional to $1/t^2$, where $t$ is the
cosmic time, may arise in certain non-critical string theory models \cite{emn}.
In such a model, the r\^ole of the quintessence field is played by the dilaton
field~\cite{aben}.  However, at least in the string theory context, a complete
treatment of such a rolling dilaton field requires a non-trivial discussion of
off-shell physics, which we present in the next section. For the moment, we
neglect this complication (as might be valid for some non-string models of
relaxing dark energy), and assume a naive on-shell physical framework in which
the dark energy density relaxes according to a generic power law $1/t^n$.
Within this general framework, $n = 0$ corresponds to a cosmological constant
$\Lambda$ and $n = 2$ corresponds to a naive interpretation of the Q-cosmology
model of Ref.~\cite{emnw}, in which only the conventional type of matter exists
and the off-shell terms are not taken into account. Such a relaxing dark energy
model would predict a Hubble parameter of the form:
\begin{equation}
H(z) = H_0 \left(\Omega_{\rm M} (1 + z)^3 + \Omega_\phi (1 + z)^n\right)^{1/2},
\label{hubbledilaton}
\end{equation}
where $\Omega_\phi$ represents the present-day dilaton dark energy in the
on-shell formulation.\footnote{However, we emphasize again that this on-shell
treatment is incorrect in the context of non-critical string theory, as we
discuss below.} In such a model with $n = 2$, the rolling dilaton dark energy
would resemble a spatial curvature contribution, and such naive on-shell models
are easily excluded on the same observational grounds that support
experimentally the spatial flatness of the Universe. However, $n = 0$ is
certainly allowed, and it is important to know the permissible range of $n$.
There is a formal correspondence between \eqref{hubbledilaton} and a
generalized $\Lambda$CDM model in which the vacuum energy $\Lambda$ obeys an
equation of state with $w_0 \ne -1$ \eqref{hubble}, namely $n =  3(1 + w_0)$.
However, the underlying physics may be entirely different. For models with
relaxing dark energy \eqref{hubbledilaton}, the standard data analysis shows
that both the `gold' and the SNLS dataset favour a value of the parameter $n$
appearing that is $\ll 2$. Indeed, the combined data prefer $n \sim 0$ ($w_0
\sim -1$), i.e., a (nearly) constant vacuum energy, as in the $\Lambda$CDM
model.

\section{Non-Critical Liouville String Q-Cosmologies}\label{sec:qcosmo}

In this Section we explore in more detail Q-cosmologies with relaxing vacuum
energy in the context of Liouville strings, which provide a consistent
treatment of off-shell, out-of-equilibrium effects. For the benefit of the
reader, we review here the basic formalism and summarize the results, giving
details of their derivation in the next two subsections, that may be skipped by
a reader not interested in technical aspects. There are many specific
cosmological models that may be formulated within the general framework of
Liouville strings: see~\cite{emnw,diamandis,diamandis2} for an extended
discussion and references. Here we limit ourselves to the main features of this
cosmological framework, which do not depend on the details of the underlying
microscopic string/brane models or compactification. We simply assume the
theory to have been compactified (somehow) to a four-dimensional
Robertson-Walker space-time, with a scale factor $a(t_E)$ in the so-called
Einstein cosmic frame \cite{aben}.

\subsection{General Formalism of Off-Shell Liouville Q-Cosmologies}

In the absence of matter, the Liouville-dressing approach of Ref.~\cite{emn}
and the (dynamical) identification of the Liouville zero mode with target time
lead to generalized conformal invariance conditions for the fields of the
gravitational multiplet of the string propagating in a four-dimensional
background:
\begin{equation}
{\ddot g}^i + Q{\dot g}^i = - \beta^i ,
\label{liouveq}
\end{equation}
where the overdot denotes differentiation with respect to the world-sheet zero
mode of the Liouville field $\phi$, the set of $g^i = \{ G_{\mu\nu}, \phi,
\dots \}$ are string background fields (metric, dilaton, \emph{etc.}), the
$\beta^i$ are the respective world-sheet renormalization-group
$\beta$-functions, and $Q$ is the square root of the central-charge deficit
that describes the departure from criticality. The negative sign in front of
$\beta^i$ on the right-hand side of \eqref{liouveq} corresponds to
supercriticality for the string, in which case the central charge exceed its
critical (conformally-invariant) value. It is in this case that the Liouville
mode has a time-like signature and can be identified with the target time
\cite{aben,emn}.

In critical strings, the vanishing of the $\beta^i$ would correspond to the
ordinary Einstein/matter equations of the low-energy field theory derived from
strings. The vanishing conditions correspond to an equilibrium situation.  In
non-critical string, however, these equations are replaced by \eqref{liouveq},
which express the restoration of conformal invariance by the Liouville mode.
The physical reasons for an initial departure from conformal invariance might
be a catastrophic cosmic event, such as the collision of two brane worlds or a
Big Bang \cite{emnw}. Such an event would cause a departure from equilibrium,
which may not be inappropriate for the early era of the Universe. Following the
identification of target time with the Liouville mode, the generalized
conformal invariance equations \eqref{liouveq} yield new dynamical equations
that describe the non-equilibrium dynamics of the Liouville strings
\cite{emn,emnw,diamandis,diamandis2}.

If one further assumes that, after compactification to four space-time
dimensions, one obtains a spatially-flat cosmological Robertson-Walker metric
background, e.g., in the context of a brane model, the equations
\eqref{liouveq} define the dynamics of the resulting `Liouville Q-cosmology',
as we term such a non-critical string cosmology.  The equations \eqref{liouveq}
\emph{are not} the ordinary equations of motion corresponding to a
four-dimensional gravitational effective action, but describe the dynamics of
an \emph{off-shell relaxation process}.

As explained in Ref.~\cite{emnw}, the inclusion of matter does not change
qualitatively this non-equilibrium, off-shell nature of such a model. Apart
from the formal addition of extra terms in the gravitational $\beta$-functions,
resulting from the coupling of matter to the gravitational field, which are
needed for general coordinate invariance in the target-space, in most models of
physical interest may be included in such a way that the main source of
non-criticality is the gravitational and moduli sector, while the matter
$\beta^i$ functions themselves are (almost) vanishing. In such a case, the
matter fields feel the presence of the off-shell Liouville terms on the
left-hand-side of \eqref{liouveq} via their gravitational couplings.

The presence of these off-shell Liouville terms, which are exclusive to
non-critical strings, results in a modification of the critical energy density
condition for the Liouville cosmology, which ensures spatial flatness. This
critical density is therefore \emph{different} from that in conventional
(on-shell) Friedman-Robertson-Walker (FRW) cosmologies.

The essential formalism is that of Ref.~\cite{aben}, in which all physically
relevant quantities should be expressed in the Einstein frame, in terms of the
Einstein cosmic time. The four-dimensional matter action (including radiation
fields) couples to the dilaton field non-trivially, in a way that is specific
to the various matter species, as a result of purely stringy properties of the
effective action \cite{gsw}. A generic $\sigma$-model-frame effective
four-dimensional action with dilaton potential $V(\phi)$, which could even
include higher string-loop corrections, has the form:
\begin{equation}
\begin{split}
S^{(4)} &= \frac{1}{2\alpha '} \int d^4 x \sqrt{-G} [e^{-\Psi(\phi)} R(G)
+ Z(\phi)(\nabla \phi)^2 + 2\alpha ' V(\phi) \dots ] -
\\ & \quad\frac{1}{16\pi}\int d^4x \sqrt{G}\frac{1}{\alpha
(\phi)}F_{\mu\nu}^2 - I_{\rm m}(\phi, G, {\rm matter})~,
\end{split}\label{matteraction}
\end{equation}
in the notation of Ref.~\cite{gasperini}, with the various factors $\Psi, Z,
\alpha$ encoding information about higher string loop corrections. Also,
$F_{\mu\nu}$ denotes the radiation field strength and $I_{\rm m}(\phi, G, {\rm
matter})$ represents matter contributions, which couple to the dilaton $\phi$
in a manner dictated by string theory, with specific scaling laws \cite{gsw}
under shifts of the dilaton field $\phi \to \phi + {\rm const}$. In cases
\cite{emnw} where only the string tree level plays a r\^ole for late times, the
various form factors simplify, e.g., $\Psi (\phi) = 2\phi, Z(\phi) = 4$,
\emph{etc.} However, for reasons of generality, here we keep the form
\eqref{matteraction}. As we shall see, there is sensitivity in the present data
to terms in the dilaton potential with higher powers of the string coupling.
When higher loop corrections are important, these factors have a complicated
form, for instance one has $e^{\Psi (\phi)} = c_0 e^{-2\phi} + c_1 +
c_2e^{2\phi} + \dots $, with various constants $c_i$, with the powers of the
square of the string coupling $g_s^2=e^{2\phi}$ counting closed string loops,
as appropriate for the gravitational multiplet.  For simplicity, in the present
work we ignore the antisymmetric-tensor four-dimensional field, which, as
discussed in Ref.~\cite{aben}, corresponds to an axion field.

Higher loop corrections may result in a modified dilaton potential for the
noncritical string theory of the form
\begin{equation}
V = 2Q^2 e^{2\phi} + {\tilde V} ~,
\label{dilfullpot}
\end{equation}
where the central charge deficit $Q^2$ encodes the microscopic non-equilibrium
physics \cite{emnw,diamandis}. In the models we are considering in this work,
$Q^2 > 0$ and the string theory is supercritical \cite{aben}.

Assuming a normal fluid form for matter or radiation, with stress tensor
$T_\mu^{\nu E} = {\rm diag}\left(-\rho, p\delta_i^j\right)$ in the Einstein
frame \cite{aben}, we obtain the following gravitational equations of motion in
our case (we work here in units $M_P^2 = 1/8\pi G_N = 1$, where $M_P$ is the
four-dimensional Planck constant) \cite{diamandis2}:
\begin{gather}
 3H^2 = \rho_m + \rho_\phi + \frac{e^{2\phi}}{2}\mathcal{J}_{\phi}~, \notag \\
2\frac{dH}{dt_E} = -\rho_m - \rho_\phi -p_m - p_\phi +
a^{-2}(t_E)\mathcal{J}_{ii}~,\quad i = 1,2,3, \notag \\
\rho_\phi \equiv \frac{1}{2}\left(2(\frac{d\phi}{dt_E})^2 + V(\phi)\right)~,
\quad p_\phi \equiv \frac{1}{2}\left(2(\frac{d\phi}{dt_E})^2 - V(\phi)\right)~,
\label{eqsmotion}\\
\frac{d^2{\phi}}{dt_E^2} + 3 H \frac{d{\phi}}{d t_E} +
\frac{1}{4} \; \frac{\partial V }{\partial \phi} + \frac{1}{2} \;( \rho_m -
3p_m )= - \frac{3 \;\mathcal{J}_{ii}}{2 \;a^2}- \frac{1}{2} e^{2\phi}
\mathcal{J}_{\phi}  \; , \notag
\end{gather}
where $t_E$ is the Einstein-frame Robertson-Walker cosmic time \cite{aben} and
the potential of the dilaton assumes the form \eqref{dilfullpot}, including
loop corrections ${\tilde V}$ \eqref{dilfullpot}. We note that the equations
\eqref{eqsmotion} differ from those holding in on-shell cosmologies by the
Liouville out-of-equilibrium contributions $\mathcal{J}$, which are exclusive
to our treatment \cite{emn,emnw,diamandis2}.

The equations \eqref{eqsmotion} need to be supplemented by the appropriate
Curci-Paffuti equation \cite{curci} expressing the renormalisability of the
$\sigma$-model fields \cite{diamandis,diamandis2}, which is also exclusive to
our non-critical string approach. This equation relates the dilaton world-sheet
$\beta$-function to the rest of the $\beta$-functions in the problem (for the
graviton, {\emph etc.}), and essentially expresses the (time-dependent) central
charge deficit in terms of the physical fields in the problem,
\begin{equation}
\frac{d \mathcal{J }_{\phi} }{d t_E} \;=\; - 6\; e^{\;-2 \phi}\;( H + \dot{\phi} ) \;
\frac{ \; \mathcal{J }_{ii}}{ \;a^2} + \dots \; ,
\label{CP}
\end{equation}
where the overdot henceforward denotes a derivative with respect to the cosmic
time $t_E$, the `$\dots$' indicate terms due to higher order in string loops,
and the values of $\mathcal{J}_{\phi}$ and $\mathcal{J}_{ii}$ are given by
\begin{align}\label{js}
\mathcal{J} _{\phi}& = e^{-2 \phi} ( \ddot{\phi} - {\dot{\phi}}^2 + Q
e^{\phi} \dot{\phi}) , \notag \\
\mathcal{J}_{ii}& = 2 a^2 \left( \ddot{\phi} + 3 H \dot{\phi} +
{\dot{\phi}}^2 +
( 1 - q ) H^2 + Q e^{\phi} ( \dot{\phi}+ H )\right), \quad i=1,\;2,\;3,\notag \\
\end{align}
with $q$ the deceleration parameter:
\begin{equation}
q(z) \equiv - \frac{\ddot{a} a}{{\dot{a}}^2} =
-\frac{dH/d t_E}{H^2(t_E)} - 1 =
\frac{(1 + z)dH/d z}{H(z)} - 1 ~.
\label{acceldef2}
\end{equation}
For completeness, we mention that the equations \eqref{eqsmotion} lead, after
standard manipulations \cite{emnw,diamandis2}, to the (non-)conservation
equation of matter in the presence of the non-equilibrium contributions:
\begin{equation}
\label{eqsmotion2}
\frac{d\rho_m}{dt_E} + 3H(\rho_m + p_m) +
\frac{1}{2}\frac{d Q}{dt_E}\frac{\partial{V(\phi)}}{\partial Q} - \frac{d\phi}{dt_E}
(\rho_m - 3p_m) = 6(H + \frac{d\phi}{dt_E})a^{-2}\mathcal{J}_{ii} . \\
\end{equation}
These equations are used below to obtain a general expression for the Hubble
parameter as a function of the redshift, which we use later in our fit to data.

\subsection{Derivation of Basic Phenomenological Formulae for Q-Cosmologies}

The above analysis contains non-linear equations that are hard to solve
analytically. A detailed numerical analysis is performed in
Ref.~\cite{diamandis2}, where we refer the interested reader for details.
Instead, for our phenomenological purposes here, we shall attempt an
approximate analytic solution at late epochs of the Universe, appropriate for
our phenomenological fits. Throughout this work we work in units where the
present scale factor is normalised to one, $a_0 = 1$.

To solve \eqref{eqsmotion2} in the various epochs of the Universe, it is
convenient first to split the energy density of matter into radiation $\rho_r$,
baryonic $\rho_b$ and dark-matter $\rho_d$ components, and to use simple
equations of state for the dilaton fluid, as supported by our theoretical model
\cite{emnw,diamandis,diamandis2}:
\begin{gather}
\rho_m = \rho_r + \rho_b + \rho_d \equiv \rho_r + \rho_M~, \nonumber \\
p_b =0, \quad p_d = w_d \rho_d, \quad p_r = \frac{1}{3}\rho_r, \quad
p_\phi = w_\phi \rho_\phi~.\label{step1}
\end{gather}
We have assumed a non-trivial equation of state $w_d$ for the dark matter
component, which as we discuss below, stems from the fact that its scaling with
the scale factor $a$ is \emph{exotic}, different from the ordinary dust scaling
$a^{-3}$. One can split the matter evolution equation \eqref{eqsmotion2} into
various components, using \eqref{step1}.

In the solution of Ref.~\cite{diamandis} without matter, at late eras the
dilaton $\phi$ and the scale factor $a(t)$ vary as follows with the cosmic time
$t_E$:
\begin{gather}
\phi = -{\rm ln}a(t_E)~, \nonumber \\
a = a_1\left( 1 + \gamma^2 t_E^2 \right)^{1/2}~, \label{dilatondiamant}
\end{gather}
where $a_1$, $\gamma$ are appropriate positive constants.\footnote{In the model
of Ref.~\cite{diamandis} they are related to the flux in the extra compactified
dimensions of the appropriate string theory, but they are viewed here as
arbitrary constants in our more general setting.} The form of the scale factor
\eqref{dilatondiamant} implies the following form for the deceleration
parameter at late epochs of the Universe:
\begin{equation}
q(z) = -1/(\gamma^2 t_E^2)~,
\label{deceltoday}
\end{equation}
The values of the constants are such that the present era in the history of the
Universe is characterised by $\gamma t_E = \mathcal{O}(1)$, as follows from the
fact that the deceleration parameter of the Universe is currently observed
\cite{wmap3} to be $q_0 = -0.61$.

We notice that the asymptotic behaviour of the model of Ref.~\cite{diamandis}
without matter, at very late times, indicates that
\begin{equation}
{\dot \phi} + H \simeq 0,
\label{phiH}
\end{equation}
and thus
\begin{equation}
H \sim a^{-1},
\label{hubble2}
\end{equation}
implying that $q \sim 1/aH $, varying very little at late epochs.

In the present case, where matter effects are important, the formulae need some
modification. In general, the dilaton and the scale factor differ from the
particular form \eqref{dilatondiamant}~\cite{diamandis2}, in which case one
obtains a much more complicated behaviour for $q(z)$ than the one given in
\eqref{deceltoday}, which can only be studied numerically. A complete numerical
analysis is performed in Ref.~\cite{diamandis2}, where we refer the reader for
details. We recall, for completeness, that the analysis of
Ref.~\cite{diamandis2} indicates evidence for a past deceleration of the
Universe at redshifts larger than $z_*=0.37$.

For our generic phenomenological purposes in this work we can make several
physically sensible simplifying assumptions, which allow us to carry some of
the results of Ref.~\cite{diamandis} through to the present case, allowing for
some analytic treatment. Specifically, the dilaton and scale factors are
assumed to take the form
\begin{equation}
\phi = -{\rm ln}a(t_E) + \mathcal{O}(g_s^2)~, \label{dilatondiamant2}
\end{equation}
as dictated by a general analysis of the dilaton equation in the
model.\footnote{We thank V.~Georgalas for an informative discussion on this
point.} Notice that, in view of the general discussion in
Ref.~\cite{diamandis2}, we no longer maintain the explicit form of $a =
a_1\left( 1 + \gamma^2 t_E^2 \right)^{1/2}$ used in the matter-less model of
Ref.~\cite{diamandis}, however we do maintain the scaling relation
\eqref{hubble2}, which, according to the numerical analysis of
Ref.~\cite{diamandis2}, seems to be a good approximation in the eras of
interest to us here.

In general \cite{diamandis2}, there may be a more complicated behaviour than
\eqref{phiH} or \eqref{hubble2}. For instance, loop and other corrections may
yield power-series corrections in the (perturbative) string coupling
\begin{equation}
{\dot \phi} + H = \mathcal{O}(H g_s^2) + \dots~, \qquad ~g_s= e^\phi ~.
\label{phiHcor}
\end{equation}
Such corrections could lead to appreciable deviations in the behaviours of
various fields and other cosmological parameters even at the present era. In
fact, the numerical analysis of Ref.~\cite{diamandis2} indicates evidence for a
past deceleration of the Universe at redshifts smaller than $z_* = 0.37$.
Nevertheless, even in these more general models, the Hubble parameter maintains
\cite{diamandis2} a scaling similar to \eqref{hubble2} at the late eras of
interest to us here, which implies that some of the qualitative features of our
analytic approach used here are valid, to a good approximation, in these more
complicated cases.

On using \eqref{eqsmotion} and performing some elementary manipulations, we
easily obtain:
\begin{equation}
\rho_M \simeq
2(2q -1)H^2 - \frac{1}{2}\frac{\partial V}{\partial \phi} +
\frac{3}{2}V + \frac{1}{2}e^{2\phi}\mathcal{J}_\phi~.
\label{eqrhoM}
\end{equation}
A consistent assumption, as we shall demonstrate now, is that the matter energy
density, $\rho_M$, includes dust (baryonic matter) and exotic scaling (dark)
components, which can behave like radiation-like contributions:
\begin{equation}
\rho_M \sim \rho_{\rm dust}^0 a^{-3} + \rho_{\rm exotic}^0 a^{-4}
+ \dots ~,
\label{darkmatterenergy}
\end{equation}
Indeed, taking into account \eqref{hubble2}, we observe that the above
assumption is consistent with the conservation equation \eqref{eqsmotion2}
provided that the central charge deficit at late eras scales as
\begin{equation}
Q^2 (a) \simeq Q_*^2 + \frac{\rho_{\rm dust}^0}{a}
\label{ccdtoday}
\end{equation}
with $Q_*^2 >0$ a constant, namely the central charge of the asymptotic
conformal field theory to which the string model flows at infinite time
\cite{emnw,diamandis}.

Notice that, if one allowed an exotic $a^{-2}$ scaling of matter of the form
$\rho_{\phi -{\rm like}}^0 a^{-2}$, which characterises the dilaton dark energy
$\rho_\phi$, as we discuss below, then $Q^2$ should contain also terms of the
form $\rho_{\phi -{\rm like}}^0 {\rm ln}(a^2)$. This would lead, in general, to
logarithmic scaling $a^{-2}{\rm ln}(a)$ of the various energy density
components, which would complicate the situation, as far as the validity of the
approximation \eqref{hubble2} is concerned. Although such terms may exist in
some models, we do not consider them in this work. Here we assume the scaling
\eqref{darkmatterenergy} for the matter energy densities in the eras of
interest to us. For more complete, numerical treatments we refer the interested
reader to Ref.~\cite{diamandis2}.

To proceed with a consistency check and estimates on the scaling behaviour of
$\rho_m,\rho_\phi$, it is necessary to make some assumptions about the form of
the potential ${\tilde V}$. As already mentioned, we assume that it is
generated by higher string loops. In brane-inspired models such as those
considered in Refs.~\cite{diamandis,emnw}, one has open strings living on the
brane world. The corrections then will be proportional to an extra power of the
string coupling $g_s = e^\phi$. Assuming loop corrections of the
form\footnote{The dots indicate higher-order corrections in the string coupling
and also possible dilaton-independent contributions.}
\begin{equation}
{\tilde V} \sim \alpha e^{3\phi} + \beta e^{4\phi}
+ \dots ~, \qquad \alpha,\beta, ... = {\rm const.} ,
\label{corr}
\end{equation}
and taking into account \eqref{ccdtoday}, we observe that the dark matter
density \eqref{eqrhoM} yields:
\begin{equation}
\rho_M \sim \frac{2(2q-1) + Q_*^2}{a^{2}}
+ \frac{\rho_{\rm dust}^0}{a^3} - \frac{\beta}{2}\frac{1}{a^4}
+ \frac{1}{2}e^{2\phi}\mathcal{J}_\phi + \dots~.
\label{rhom2}
\end{equation}
As we see from \eqref{phiHcor}, \eqref{hubble2}, \eqref{js}, \eqref{ccdtoday},
the quantity $\mathcal{J}_\phi = \mathcal{O}(a^{-2}) $ is negative. We also
note that the late-era scaling \eqref{darkmatterenergy}, which is required for
consistency (on account of \eqref{ccdtoday}), can easily be guaranteed,
provided the terms $2(2q-1)+ Q_*^2$ which otherwise would yield a $a^{-2}$
scaling are suppressed for the redshift region $0 < z < 2$ of interest to us.
This would require a $Q_*^2$ of order $2|(2q_0 -1)| $, according to our
assumption that $q$ does not change much in the region in which we are
interested.\footnote{We remark at this point that the actual data may deviate
from this for redshifts less than $0.5$, indicating past deceleration of the
Universe. However, even in such a case, the absence of $a^{-2}$ terms in matter
today might be a consistent assumption. In fact, the numerical solution of
Ref.~\cite{diamandis2} indicates that the matter does develop an exotic scaling
$a^{-2}$ but at much later eras than the present one.}

Notice that the dust-like scaling does not receive contributions from the
string loop corrections, and in fact a comparison between \eqref{rhom2} and
\eqref{eqrhoM} yields a mathematically consistent result with $\rho_{\rm
dust}^0$  free parameter to be constrained by fitting data and/or other
microscopic model considerations.

In contrast, the exotic scaling $a^{-4}$ receives contributions from the loop
corrections, proportional to $\beta$, whose sign cannot be fixed by our generic
considerations, and depends on the details of the microscopic string/brane
theory model, and contributions due to the off-shell Liouville terms
$e^{2\phi}\mathcal{J}_\phi$ which are positive according to our assumption that
$Q_* > 0$:
\begin{equation}
\rho_{\rm exotic}^0 \sim -\frac{\beta}{2} - \frac{|q| + Q_*}{2}.
\label{exotictoday}
\end{equation}
The asymptotic behaviour of the dilaton dark energy may be evaluated from the
definition of $\rho_\phi$ \eqref{eqsmotion}:
\begin{equation}\label{rhophi}
\begin{split}
\rho_\phi& = {\dot \phi}^2 + \frac{V}{2} \sim H^2 + \frac{Q_*^2}{a^2} +
\frac{\rho_{\rm dust}^0 + \alpha/2 }{a^3} + \frac{\beta }{2~a^4}
+ \dots \\
& = \mathcal{O}(a^{-2})  + \mathcal{O}(a^{-3}) + \frac{\beta}{2}~a^{-4} +
\dots ,
\end{split}
\end{equation}
where the dots have the same meaning as in \eqref{corr}. It is interesting to
observe that the dilaton dark energy component contains a part scaling like
$a^{-2}$, which is positive and of order $(1 + Q_*^2)a^{-2}$, but also parts
scaling like $a^{-3}$ and $a^{-4}$, which appear also in the matter energy
density \eqref{eqrhoM}. However, in contrast to the matter case \eqref{rhom2},
the `dust'-like contributions to the dilaton dark energy do not have a fixed
sign or magnitude, as they depend on the string loop correction parameter
$\alpha$,
\begin{equation}
\rho^0_{\phi , 3} \sim 2\rho_{\rm dust}^0 + \alpha/2 \sim (1 + \alpha )/2 .
\label{dustinphi}
\end{equation}
When combined with the corresponding parts in $\rho_M$, in order to obtain the
precise scaling of the Hubble parameter with the redshift $z$, $H(z)$, from the
first of equations \eqref{eqsmotion}, we observe that
\begin{equation}
\rho_\phi + \rho_M \simeq |\mathcal{O}(a^{-2}) | +
\frac{4\rho_{\rm dust}^0 + \alpha}{2}a^{-3} + \left( -\frac{|q| + Q_*}{2}
+ \dots \right) a^{-4},
\label{predictions}
\end{equation}
where we have been careful to indicate the relative signs of the various terms,
as predicted by the model with the approximations made so far.\footnote{One may
have deviations from \eqref{phiHcor} \cite{diamandis2}, resulting in extra
contributions to the $a^{-4}$ scaling, denoted by dots. Thus, the exotic (dark)
matter $\Omega_\delta$ could be positive. This case seems to be favoured by the
data, as discussed in the text, and it is in agreement with the numerical
analysis of Ref.~\cite{diamandis2}.} Notice the independence of the exotic
scaling on the loop parameter $\beta$, which does not enter the expression for
$H(z)$. Thus, it is a feature of this Liouville model that \emph{the sign of
the dust-like contribution appearing in $H(z)$ is not fixed}, since such terms
receive contributions from the dilaton dark energy. Indeed, in modern
string/membrane models, one may obtain negative dust-like contributions to the
dilaton potential from compactification of Kaluza-Klein graviton modes in brane
models, \emph{etc.} \cite{kaluza}. In contrast, the exotic scaling terms appear
positive.

When fitting the data, in order to distinguish ordinary matter from dilaton
dark energy effects that scale similarly with the redshift, one would have to
perform also measurements of the equation of state of each component, which we
do not attempt in this work. In the specific model \cite{diamandis} considered
here as our pilot study, the values of the various $\Omega_i$ appearing in the
expression for $H(z)$ are in principle predicted by the underlying microscopic
dynamics of the model, as we have seen above, but we leave this exercise for
future work.

However, since some of the $\Omega_i$ involve highly model-dependent
coefficients of string loop corrections and other, currently unknown, details
of microscopic models, and, moreover, in the above derivation several
assumptions have been made, e.g., about the form of the potential for the
dilaton, the form of the dilaton and the scale factor at late epochs
\eqref{dilatondiamant}, \emph{etc.}, which may not always be valid
\cite{diamandis2}, we consider at present the various parameters appearing in
the expression for $H(z)$ as arbitrary parameters to be fixed by the data. To
be even more general, we keep the exponent of the exotic scaling as a free
fitting parameter as well. From the above considerations, in particular
\eqref{darkmatterenergy}, \eqref{rhophi}, it is also clear that in general
$\Omega_{\rm dust}$ may be negative, as its sign depends on the sign of the
corrections \eqref{corr} to the dilaton potential.

The final result of our parametrisation for $H(z)$ in the Q-cosmology
framework, which will be compared with the astrophysical data in the next
section, in order to obtain information on the basic cosmological parameters of
the model, is therefore:
\begin{equation}\label{formulaforfit}
H(z) = H_0 \left( {\Omega }_3 (1 + z)^3 + {\Omega }_{\delta} (1 +
z)^\delta + {\Omega}_2(1 + z)^2 \right)^{1/2}~,
\end{equation}
with the densities $\Omega_{2,3,\delta}$ corresponding to present-day values
($z = 0$) and
\begin{equation}
\label{sumomegas}
{\Omega }_3 + {\Omega}_{\delta} + {\Omega}_2 = 1~,
\end{equation}
The reader should recall that in the model described in this section $\delta =
4$. However, in view of the various approximations employed in our analytic
treatment of $Q$-cosmology above, which, as already mentioned, may not be valid
in the present era, when matter contributions are important, the exponent
$\delta$ is treated from now on as a fitting parameter. The above formulae are
valid for late eras, such as the ones pertinent to the supernova and other data
($0 \le z \le 2$) that we use in this work.

We stress once again that the various $\Omega_i$ contain contributions from
\emph{both} dark energy and matter energy densities. As explained previously,
$\Omega_3$ does not merely represent ordinary matter effects, but also receives
contributions from the dilaton dark energy. In fact, the sign of $\Omega_3$
depends on details of the underlying theory, and it could even be
\emph{negative}. For instance, Kaluza-Klein graviton modes in certain
brane-inspired models \cite{kaluza} yield negative dust contributions. In a
similar vein, the exotic contributions scaling as $(1+z)^\delta$ are affected
by the off-shell Liouville terms of Q-cosmology. It is because of the similar
scaling behaviours of dark matter and dilaton dark energy that we reverted to
the notation $\Omega_i$, $i=2,3,\delta$ in \eqref{formulaforfit}. To
disentangle the ordinary matter and dilaton contributions one may have to
resort to further studies on the equation of state of the various components,
which we do not study in this article. More generally, one could have included
a cosmological constant $\Omega_\Lambda$ contribution in \eqref{formulaforfit},
which may be induced in certain brane-world inspired models. We do not do so in
this work, as our primary interest is to fit Q-cosmology models
\cite{emnw,diamandis,diamandis2}, which are characterised by dark energy
densities that relax to zero.

\section{Data Analysis}\label{sec:data}

\subsection{Astrophysical Observables}

We now discuss the observables available to test the above models using mainly
data on high-redshift supernovae and (to a limited extent) baryon acoustic
oscillations.

We use the supernovae data reported in Refs.~\cite{deceldata,SNLS}, which are
given in terms of the distance modulus
\begin{equation}
\mu = 5\log d_L + 25~.
\label{modulus}
\end{equation}
We use for $d_L$ the standard formula relating the luminosity distance to the
redshift $z$
\begin{equation}
d_L = c(1 + z) \int_0^z \frac{dz'}{H(z')} ,
\label{luminositydistanceredshift}
\end{equation}
where $H$ is given by \eqref{hubble},\eqref{modhubble} and
\eqref{formulaforfit} in the three models discussed in the previous sections,
namely $\Lambda$CDM, super-horizon and Q-cosmology respectively. We note that
this observable depends on the expansion history of the Universe from $z$ to
the present epoch, and recall that most of the available supernovae have $z <
1$, though there is a handful with larger values of $z$.

For standard on-shell cosmologies, complementary information is provided by the
data on baryon acoustic oscillations (BAO) \cite{linder,baryon}, which show up
in the galaxy-galaxy correlation function at $z \sim 0.35$. However, for
reasons to be discussed later on in the article, the application of such an
analysis to the off-shell Q-cosmology model is an open issue, since the
underlying theoretical framework needs to be re-evaluated.

The formulae for $H(z)$ in the various models discussed above, namely
\eqref{hubble}, \eqref{modhubble} and \eqref{formulaforfit}, serve as the basis
for our analysis below.

\subsection{Experimental Fits}

We now present our fits of the above models to the presently available
astrophysical data on high-redshift supernovae \cite{deceldata,SNLS}, and (in
the case of $\Lambda$CDM) also baryon acoustic oscillations (BAO)
\cite{linder,baryon}. We start our discussion with the $\Lambda$CDM and
Super-Hubble-Horizon models, and then proceed to the case of off-shell
Liouville Q-cosmology models \cite{emnw,diamandis,diamandis2}.

\begin{figure}[htb]
\begin{center}
\includegraphics[width=0.7\linewidth]{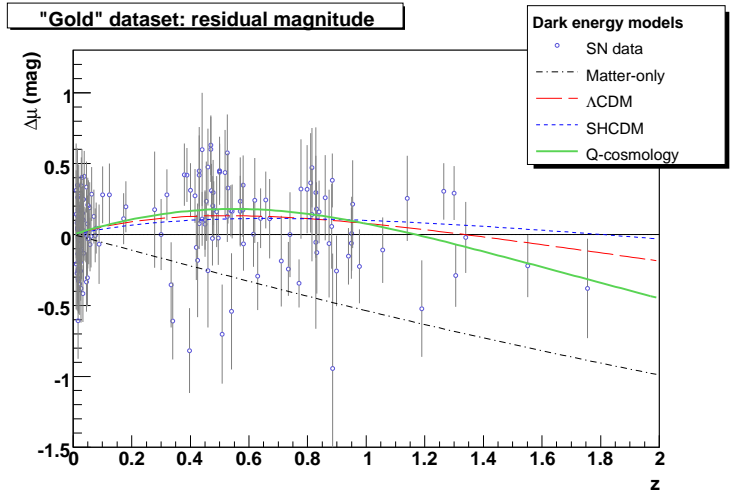}
\end{center}
\caption{Residual magnitude of the `gold' dataset supernovae as a function of
the redshift for the best-fit values of the cosmological parameters. The
following models are shown: (i) Empty Universe (black solid line); (ii)
Universe with matter only, $\Omega_{\rm M}=1$ (black dashed-dotted line); (iii)
$\Lambda$CDM model for $\Omega_{\rm M} = 0.287$ (red long-dashed line); (iv)
the super-horizon model for $\Psi_{\ell0} = -0.79$ (blue short-dashed line);
and (v) off-shell Q-cosmology model for $\Omega_3=-3.6$ and $\Omega_4=1.2$
(green thick solid line). } \label{fig:gold_datafit}
\end{figure}

\begin{figure}[htb]
\begin{center}
\includegraphics[width=0.7\linewidth]{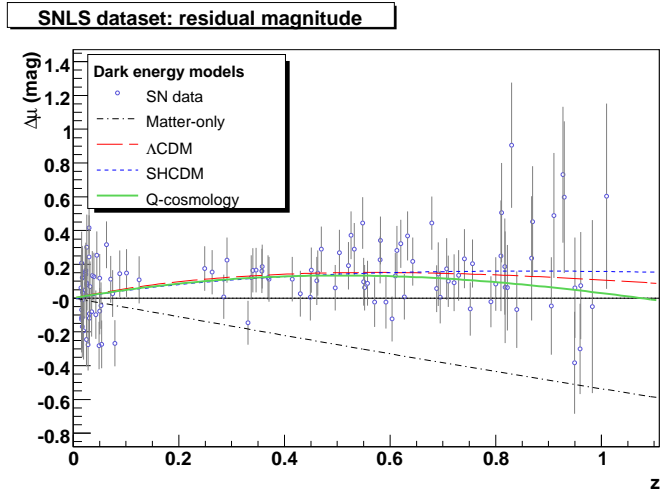}
\end{center}
\caption{Residual magnitude of the SNLS dataset supernovae as a function of the
redshift for the best-fit values of the cosmological parameters. The following
models are shown: (i) Empty Universe (black solid line); (ii) Universe with
matter only $\Omega_{\rm M}=1$ (black dashed-dotted line); (iii) $\Lambda$CDM
model for $\Omega_{\rm M} = 0.265$ (red long-dashed line); (iv) the
super-horizon model for $\Psi_{\ell0} = -0.94$ (blue short-dashed line); and
(v) off-shell Q-cosmology model for $\Omega_3=-3.0$ and $\Omega_4=1.0$ (green
thick solid line).} \label{fig:snls_datafit}
\end{figure}

Measurements \cite{snIa,deceldata} are available of the distance moduli
\eqref{modulus} for 157~supernovae in a so-called `gold' sample
\cite{deceldata}, observed by ground-based facilities and the Hubble Space
Telescope: see Fig.~\ref{fig:gold_datafit}. We compare these with the
measurements of 71~other high-redshift supernovae published more recently
\cite{SNLS} by the Supernova Legacy Survey (SNLS), which are accompanied by a
reference sample of 44 nearby SN~Ia, yielding a total of 115~data points: see
Fig.~\ref{fig:snls_datafit}. For both samples, the data are expressed in terms
of the observed distance modulus $\mu_{ob}$ derived from the SN~Ia light curves
and the corresponding redshift $z$. The best-fit cosmological parameters are
then extracted by minimizing:
\begin{equation}\label{eq:chi2}
\chi^2 =
\sum_{i}\frac{(\mu_{ob,i}-\mu_{th}(z_i))^2}{\sigma_{ob,i}^2+\sigma_{int}^2},
\end{equation}
where $\mu_{ob,i}$ and $\sigma_{ob,i}$ are the measured value for the distance
modulus and the corresponding error for a specific supernova $i$, respectively,
$\sigma_{int}$ is the intrinsic dispersion of the absolute magnitudes of the
supernovae, and $\mu_{th}(z_i)$ is the theoretical prediction for a given model
as calculated by \eqref{modulus}, \eqref{luminositydistanceredshift}, where
$d_L$ is given in units of megaparsecs. This fit and all subsequent analyses
are performed with the ROOT \cite{root} implementation of the Minuit function
minimization and error analysis code \cite{minuit}.

Both the data and the predictions of the various models are expressed in the
following as residuals, $\Delta\mu$, from their predicted values in an empty
Universe ($\Omega_{\rm M}=0$):
\begin{equation}\label{eq:empty}
d_L^{(empty)} = \frac{c}{2H_0}z(2+z).
\end{equation}
In addition to presenting the `gold' and the SNLS data sets respectively,
Figures~\ref{fig:gold_datafit} and \ref{fig:snls_datafit} also display the
predictions of various cosmological models to be detailed later on in the
article.

In addition to the 157~SN~Ia belonging to the `gold' sample analysed in
Ref.~\cite{deceldata}, a so-called `silver' dataset of 29~SN~1a is also
available, with lesser spectrometric and photometric quality compared to the
`gold' one. Although the results presented here were obtained with the `gold'
dataset, the analysis was also repeated including the `silver' supernovae data
---with a total of 186~supernovae--- with results comparable to those presented
here, proving the robustness of the analysis. The analysis was also
repeated for the combined sample of the `gold' (157~supernovae) plus the
SNLS sample (71~supernovae), which yields a sample of 228~supernovae.

The $\chi^2$ values obtained for the best fits to the various models are
discussed in the remainder of this Section.

\subsubsection{Cosmological-constant model}

\begin{figure}[tb]
\begin{center}
\includegraphics[width=0.4\linewidth]{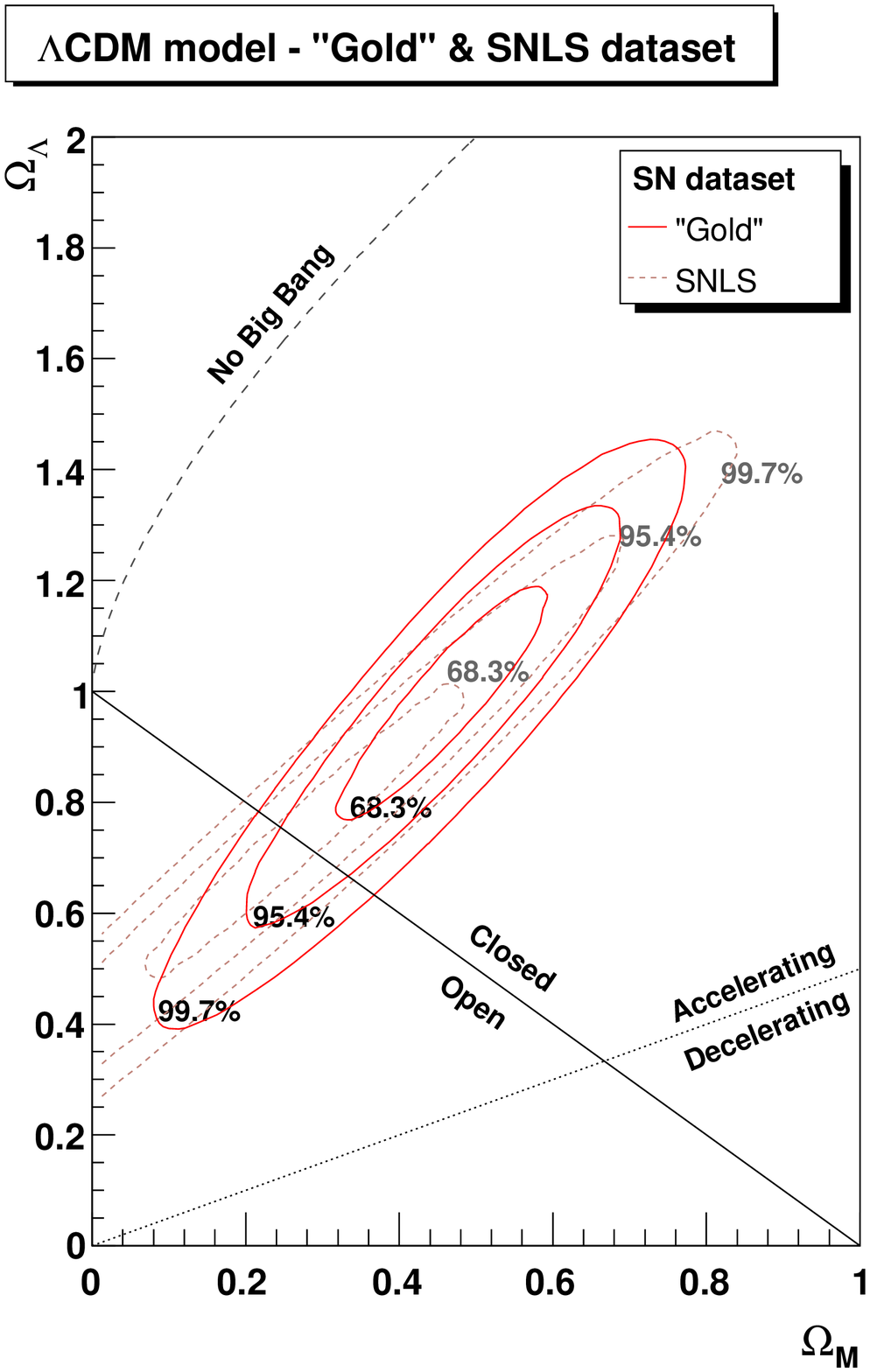}
\includegraphics[width=0.4\linewidth]{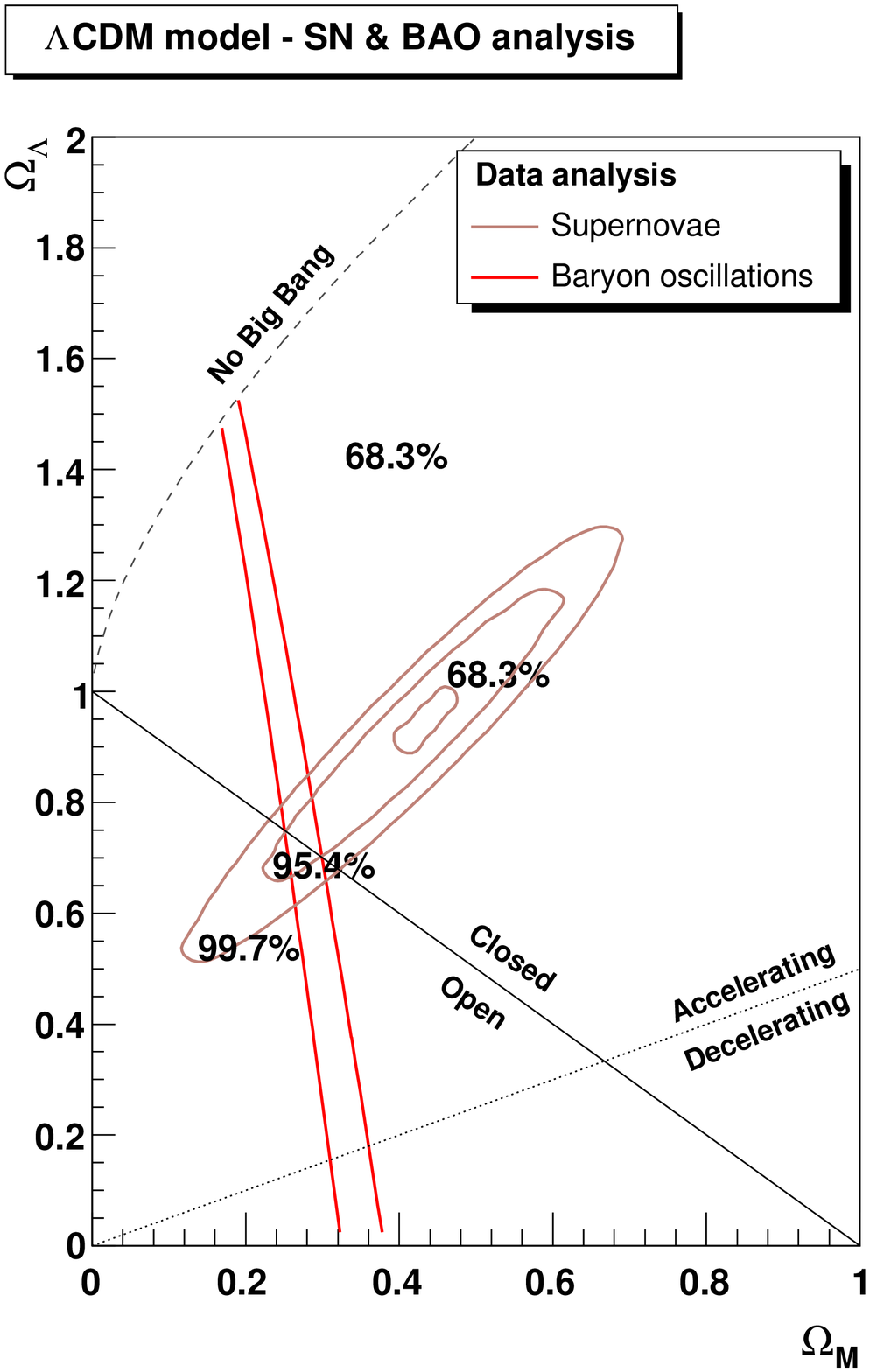}
\end{center}
\caption{Confidence-level contour plots for the $\Lambda$CDM model. (i) Left
panel: `gold' sample of 157~supernovae \cite{deceldata} and SNLS sample of
71+44~supernovae \cite{SNLS}, (ii) right panel: combined `gold' and SNLS sample
of 157+71~supernovae. The 68.3\% C.L.\ region extracted from the BAO analysis
is also superimposed.} \label{fig:lcdm}
\end{figure}

For the $\Lambda$CDM model \cite{concordance}, assuming a flat Universe, we
find the $\chi^2$ and parameter values listed in Table~\ref{tb:lcdm}, which are
in agreement with those presented in Ref.~\cite{deceldata,SNLS}. The
corresponding confidence limits in the $(\Omega_{\rm M},\Omega_{\Lambda})$
plane are shown in Fig.~\ref{fig:lcdm}.

\begin{table*}[H]
\caption{Fits to the $\Lambda$CDM model parameter $\Omega_{\rm M}$, assuming a
flat Universe. We compare the values favored by the `gold' and the SNLS data
sets, and give in the third row the results of a fit to the combined data set.}
\label{tb:lcdm}
\begin{center}
\begin{tabular}{ l l l l l } \hline
  SN data set & $\Omega_{\rm M}$ & $\chi^2$ & $\chi^2/{\rm dof}$ \\ \hline
  `gold'      & $0.287\pm0.026$  & 178      & 1.14 \\
   SNLS       & $0.265\pm0.022$  & 114      & 1.00 \\
   Combined   & $0.274\pm0.017$  & 239      & 1.05 \\
  \hline
\end{tabular}
\end{center}
\end{table*}

We see from Table~\ref{tb:lcdm} and Fig.~\ref{fig:lcdm} that the fits to the
two data sets are quite compatible. It is therefore reasonable to make a joint
fit to the combined data set, which yields the results shown in the third row
of Table~\ref{tb:lcdm}: the low value of the $\chi^2$/dof justifies this
combination {\it a posteriori}. For the value of $\Omega_{\rm M}$ found there,
the critical redshift $z_*$ above which deceleration occurs is $z_* \simeq
0.74$.

\subsubsection{Super-horizon model}

\begin{figure}[tb]
\begin{center}
\includegraphics[width=0.43\linewidth]{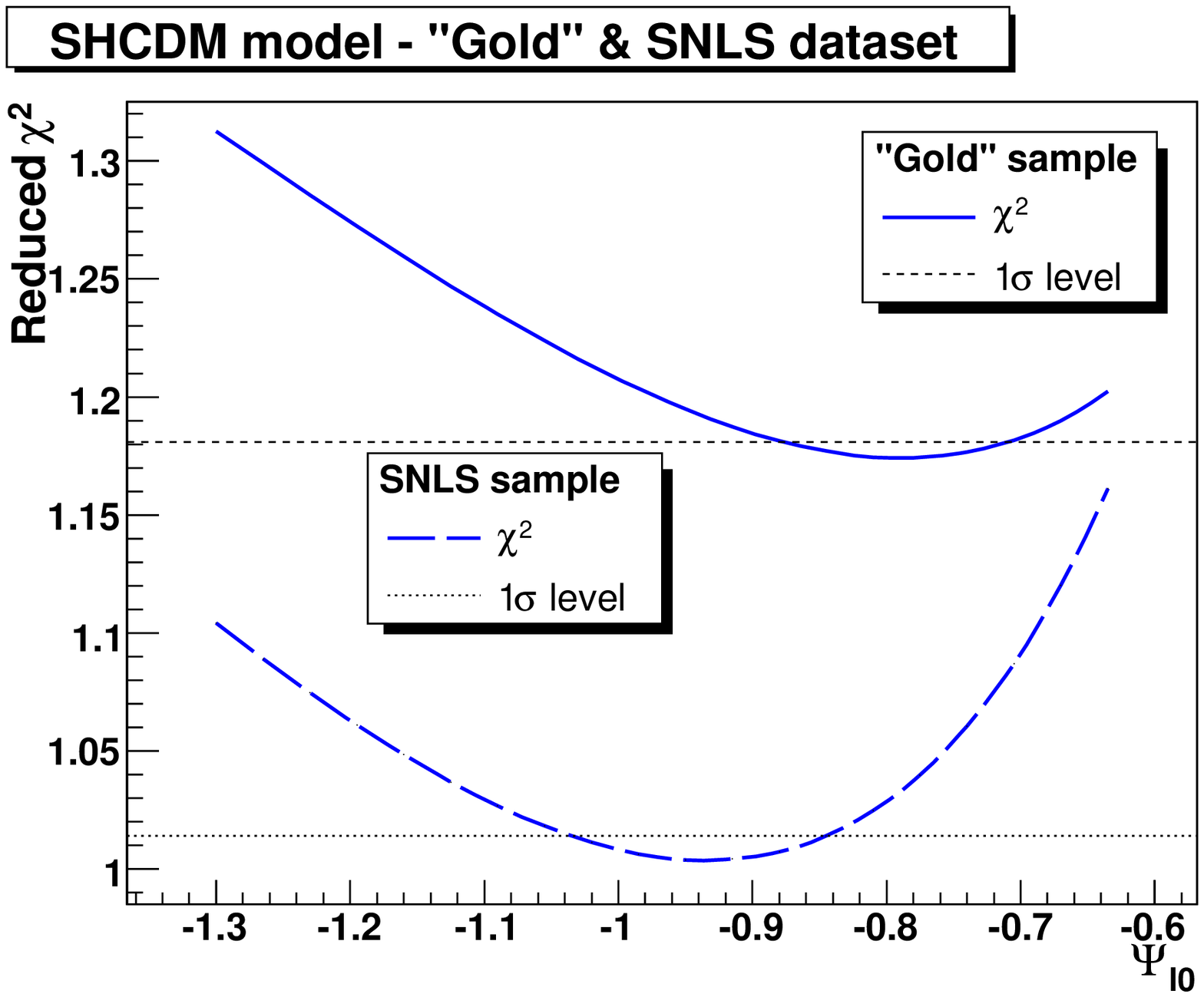}
\includegraphics[width=0.43\linewidth]{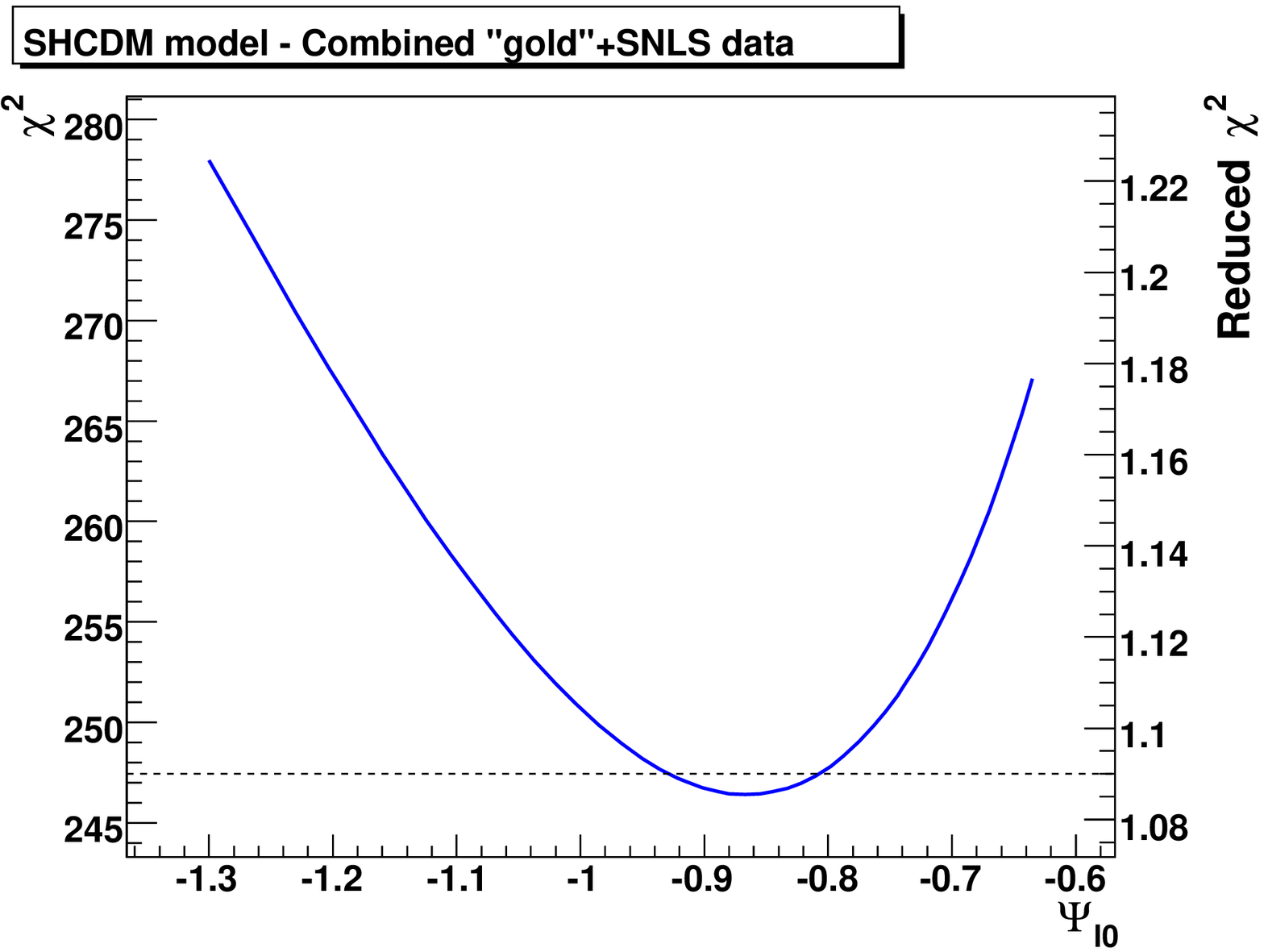}
\end{center}
\caption{The likelihood function $\chi^2$ and the reduced $\chi^2$ per degree
of freedom as a function of $\Psi_{\ell0}$ of the super-horizon model
\cite{riotto}. (i) Left figure: `Gold' sample of 157~supernovae
\cite{deceldata} and SNLS sample of 71+44~supernovae \cite{SNLS}; (ii) right
figure: combined `gold' and SNLS sample of 157+71~supernovae.}
\label{fig:shcdm}
\end{figure}

For the super-horizon model of Ref.~\cite{riotto} we find the fit values shown
in Table~\ref{SN:SHCDM}. Figure~\ref{fig:shcdm} displays the $\chi^2$ values
for the two data sets, as functions of the model parameter $\Psi_{\ell 0}$. In
this case, the consistency between the fits to the `gold' and SNLS data sets is
not as good as for the $\Lambda$CDM model. Nevertheless, the overall fit
quality is acceptable, and the third line of Table~\ref{SN:SHCDM} displays the
parameter values found in a combined fit: the low value of the $\chi^2$/dof
again justifies this combination \emph{a posteriori}. In this model, the
critical redshift is $z_* = 4.6$.

\begin{table*}[H]
\caption{Fits to the SHCDM model~\protect\cite{riotto} parameter $\Psi_{\ell0}$
using the `gold' and the SNLS data sets and, in the third line, the result of a
fit to the combined data set.} \label{SN:SHCDM}
\begin{center}
\begin{tabular}{ l l l l l } \hline
  SN data set & $\Psi_{\ell0}$ & $\chi^2$ & $\chi^2/{\rm dof}$ \\ \hline
  `gold' & $-0.79\pm0.08$ & 183 & 1.17 \\
  SNLS   & $-0.94\pm0.09$ & 114 & 1.00 \\
  Combined   & $-0.87\pm0.06$ & 245 & 1.09 \\
  \hline
\end{tabular}
\end{center}
\end{table*}

\begin{figure}[tb]
\begin{center}
\includegraphics[width=0.48\linewidth]{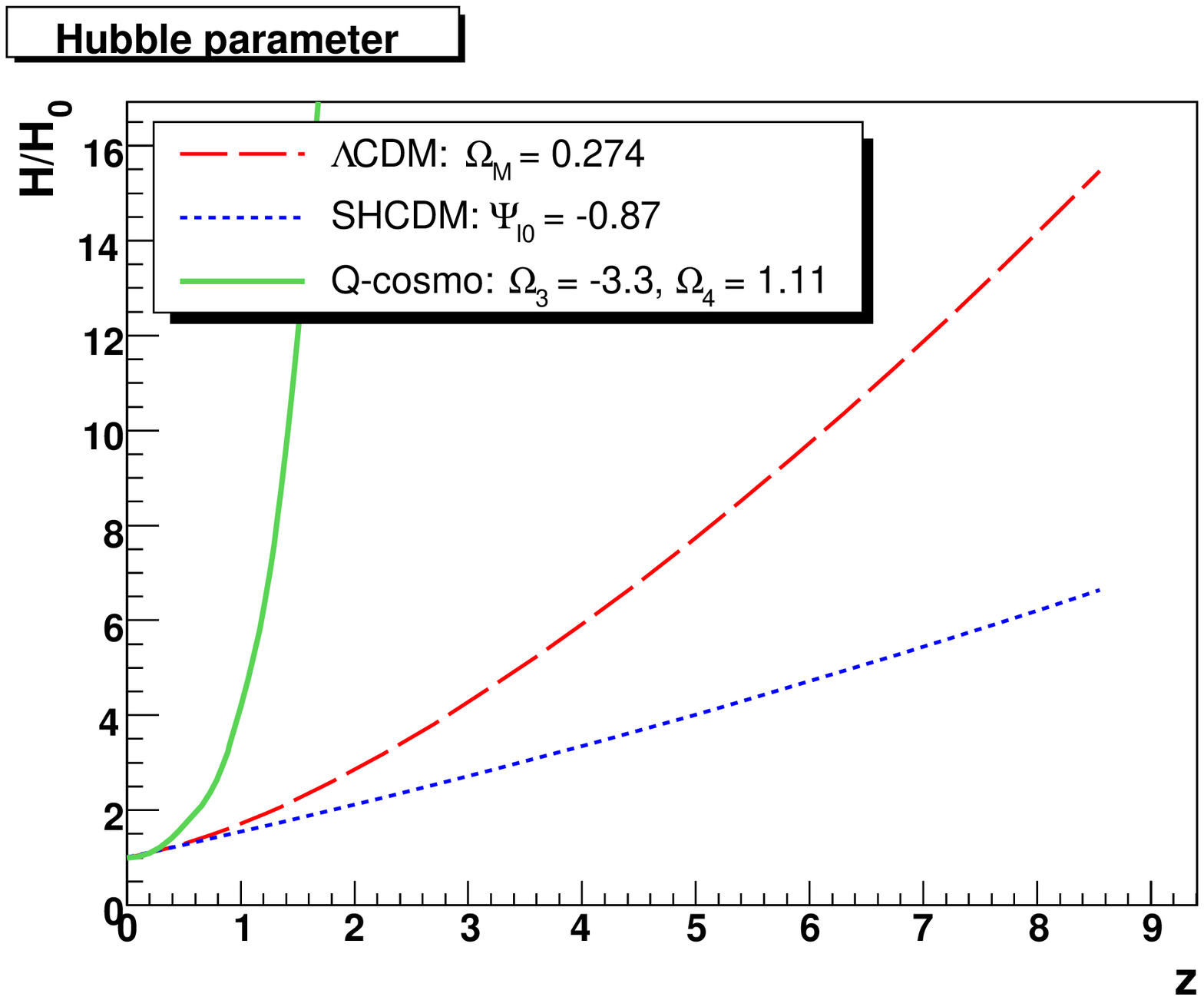}
\includegraphics[width=0.48\linewidth]{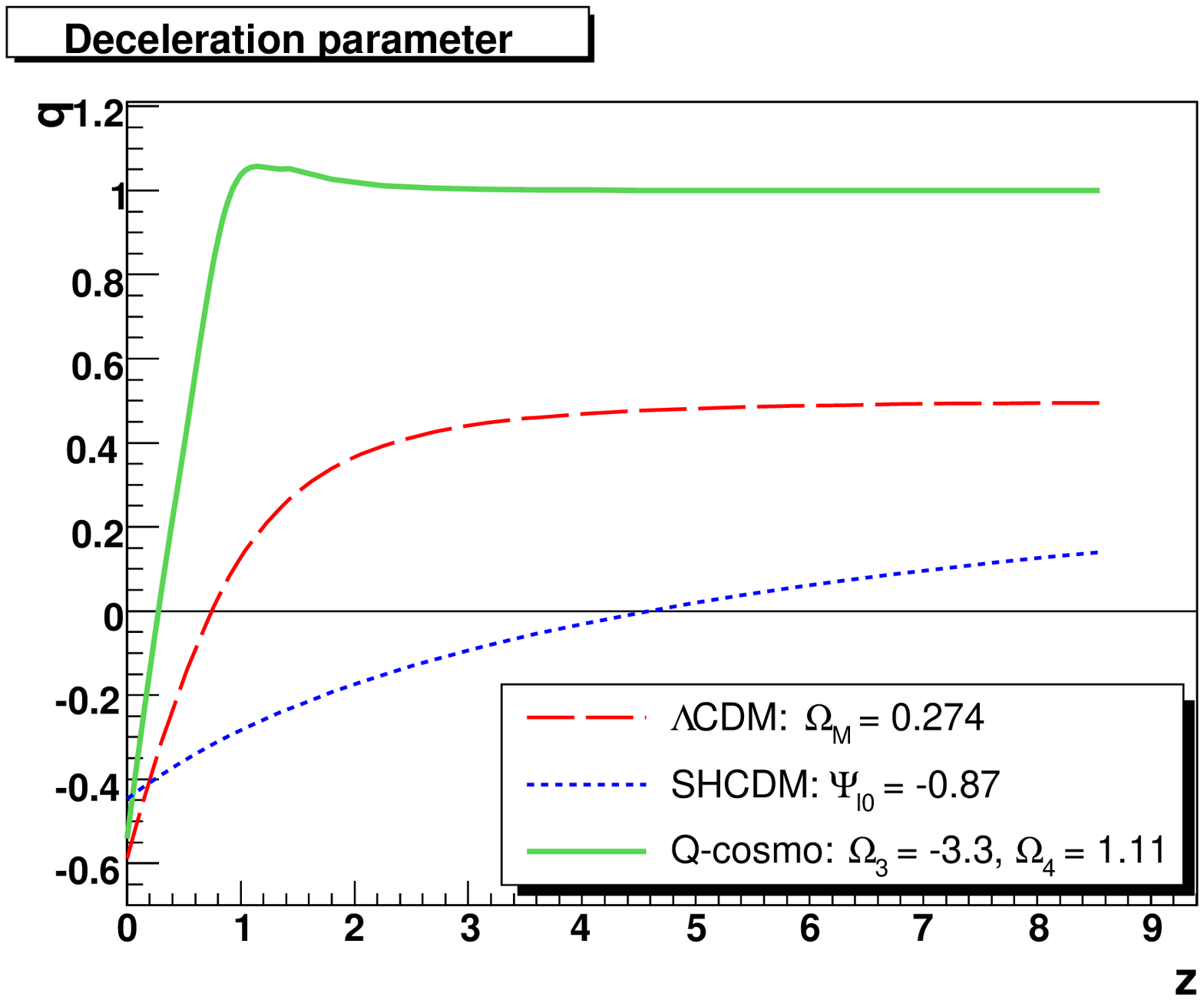}
\end{center}
\caption{The Hubble parameter (left) and the deceleration parameter (right) as
a function of the redshift $z$ for various cosmological models: (i) red long
dashed curve, the standard $\Lambda$CDM model, (ii) blue short dashed curve,
the SHCDM model of Ref.~\cite{riotto}, and (iii) green solid curve, Q-cosmology
(off-shell) model with dark energy relaxing as $1/t^2$. }\label{fig:deceler}
\end{figure}

The value of the critical redshift $z_*$ marking the transition from
deceleration to acceleration may be compared in each of the above cases with
that reported in Ref.~\cite{deceldata}.  We note that the precise value of
$z_*$ depends crucially on the microscopic model used, in particular on the
equation of state, which is in general $z$-dependent. This affects the precise
functional dependence on $z$ of the Hubble parameter $H(z)$ and the
deceleration parameter $q(z)$, as seen in Fig.~\ref{fig:deceler}. The analysis
of Ref.~\cite{deceldata}, which indicated a critical $z_* = 0.46 \pm 0.13$, was
based on a very simple assumed form (linear) for the $z$-dependent deceleration
rate, namely $q(z) = q_0 + q_1 z $, which is not the case in all models, as
seen in the right panel of Fig.~\ref{fig:deceler}.  We therefore consider that
the issue of the exact value of $z_*$ is a delicate point to be explored in the
future.

\subsubsection{Q-Cosmology (off-shell) Model}

We now present a detailed fit of the off-shell Q-cosmology model described
above to the presently available astrophysical data on supernovae
\cite{deceldata,SNLS}, checking the consistency of the model and determining
the crucial cosmological parameters of this generic class of Liouville
Q-cosmology models of Refs.~\cite{emnw,diamandis,diamandis2}.

We again analyze the supernovae data \cite{deceldata,SNLS}, as given in terms
of the distance modulus. We use the formula \eqref{luminositydistanceredshift},
with the Hubble parameter given by \eqref{formulaforfit}, to obtain $d_L$ as
predicted by the model of Refs.~\cite{emnw,diamandis}. As before, we use the
157~data points from the so-called `gold' sample of supernovae
\cite{snIa,deceldata} observed by ground-based facilities and Hubble Space
Telescope, as well as the more recent SNLS sample of 115~supernonae reported in
Ref.~\cite{SNLS}. The best-fit parameter values and the corresponding values
for the likelihood function $\chi^2$ are listed in Tables~\ref{tb:qcosmodelta}
and~\ref{tb:qcosmo4} for both datasets and their combination.

The analysis in Table~\ref{tb:qcosmodelta} has three free parameters, namely
$\Omega_3$, $\Omega_{\delta}$ and the exponent $\delta$. As in previous fits,
the parameter values favoured by the `gold' and SNLS supernovae samples are
quite compatible, so we concentrate on fit to the combined data. We see that
the three contributions to the energy density, $\Omega_{2,3,\delta}$, are not
very well determined individually, but the fit prefers a value close to 4 for
the exponent $\delta$. We therefore repeated the fit, this time fixing $\delta
= 4$, with the results shown in Table~\ref{tb:qcosmo4}. The quality of the
combined fit is no worse than with $\delta$ left free, and the values of the
different energy densities $\Omega_{2,3,4}$ are compatible with those in
Table~\ref{tb:qcosmodelta}, but with significantly smaller errors. The
Q-cosmology curves in Figs.~\ref{fig:gold_datafit} and \ref{fig:snls_datafit}
are based on the best-fit values in Table~\ref{tb:qcosmo4}.\footnote{This
analysis was repeated using also the so-called `silver' supernovae, and the
results were comparable to the ones presented here, confirming the robustness
of our analysis.} Off-shell contributions to the `dust' term $\Omega_3$ allow
it to take negative values, so the negative signs found in the fits in
Tables~\ref{tb:qcosmodelta}, \ref{tb:qcosmo4} are quite consistent with this
framework.

\begin{table*}[H]
\caption{Q-cosmology model parameter values favoured by the `gold' and the SNLS
supernovae datasets. $\Omega_2$ is determined by the other densities so that
$\Omega_{\delta}+\Omega_3+\Omega_2=1$.} \label{tb:qcosmodelta}
\begin{center}
\begin{tabular}{ l l l l l l l } \hline
  SN dataset &  $\Omega_3$ & $\Omega_{\delta}$ & $\delta$ & $\Omega_2$ & $\chi^2$ & $\chi^2/{\rm dof}$ \\
  \hline
  `Gold'   & $-3.2\pm0.4$ & $1.0\pm1.7$ & $4.10\pm0.20$ & $3.2\pm1.7$ & 177 & 1.14 \\
     SNLS    & $-2.0\pm0.8$ & $0.4\pm0.4$ &  $4.7\pm1.0$  & $2.6\pm0.9$ & 113 & 1.01 \\
  Combined & $-3.7\pm1.1$ & $1.3\pm0.7$ &  $3.9\pm0.3$  & $3.4\pm1.3$ & 237 & 1.05 \\
  \hline
\end{tabular}
\end{center}
\end{table*}

\begin{table*}[H]
\caption{Q-cosmology model parameter values favoured by the `gold' and the SNLS
supernovae datasets for a fixed value $\delta=4$. $\Omega_2$ is determined by
the other densities so that $\Omega_{4}+\Omega_3+\Omega_2=1$.}
\label{tb:qcosmo4}
\begin{center}
\begin{tabular}{ l l l l l l } \hline
  SN dataset &  $\Omega_3$ & $\Omega_4$ & $\Omega_2$ & $\chi^2$ & $\chi^2/{\rm dof}$ \\ \hline
    `Gold'   & $-3.6\pm0.8$ &  $1.2\pm0.3$  & $3.4\pm0.9$ & 177 & 1.14 \\
     SNLS    & $-3.0\pm0.9$ &  $1.0\pm0.4$  & $3.0\pm1.0$ & 113 & 1.00 \\
  Combined & $-3.3\pm0.6$ & $1.11\pm0.25$ & $3.2\pm0.7$ & 237 & 1.05 \\
  \hline
\end{tabular}
\end{center}
\end{table*}

We observe from Fig.~\ref{fig:deceler} that, with the best-fit values of the
Q-cosmology parameters in the model of Refs.~\cite{emnw,diamandis2}, the
transition from a decelerating expansion of the Universe to an accelerating
expansion would have taken place when $z = z_* \simeq 0.31$.  This is almost
compatible at the 1-$\sigma$ level with the result of the analysis of
Ref.~\cite{deceldata}, which indicated a critical $z_* \simeq 0.46 \pm 0.13$.
However, this was based on a very simple (linear) assumed form for the
deceleration $q(z) = q_0 + q_1 z$, which is not the case of our model, as also
seen in Fig.~\ref{fig:deceler}. It should be possible in the future to explore
observationally nonlinear forms of $w_\phi (z)$, with the aim of improving the
fit and reducing the uncertainty in the value of the critical $z = z_*$ for the
transition from deceleration to acceleration.

\section{Results and Outlook}\label{sec:summary}

We have shown in this paper that the two main data sets for high-redshift
supernovae are compatible, at least as far as their constraints on the
cosmological models studied in this paper are concerned. The standard
$\Lambda$CDM model fits the supernova data \emph{very well}, and the
super-horizon dark matter model also fits the supernova data \emph{quite well}.
Both of these models are on-shell, i.e., they satisfy the pertinent Einstein's
equations. However, off-shell cosmology models can still be compatible with the
data. As we discussed above, off-shell effects are important in our Q-cosmology
model. Introducing an extra parameter to allow for these off-shell effects, we
find that \emph{the Q-cosmology model may fit the supernovae data as well as
the standard $\varLambda$CDM model}.

The best fits acquired by combining the `gold' sample \cite{deceldata} and the
SNLS \cite{SNLS} supernovae data set have the following $\chi^2$ values:

\begin{itemize}

\item In the $\Lambda$CDM model that combines a cosmological constant with
cold dark matter, assuming a flat Universe we find $\chi^2=239$ ($\chi^2/{\rm
dof} =1.05$) for $\Omega_{\rm M}=0.274\pm0.017$, corresponding to a value of
$\Omega_\Lambda = 0.726 \pm 0.017$ for the dark energy density. This value is
consistent with the one predicted in Ref.~\cite{deceldata}.

\item In the super-horizon dark matter model (SHCDM) of Ref.~\cite{riotto},
we find $\chi^2 = 245$ ($\chi^2/{\rm dof} =1.09$) for the best-fit case with
$\Psi_{\ell0} = -0.87\pm0.06~.$

\item For the Q-cosmology model of Refs.~\cite{emnw,diamandis2}, assuming $\delta=4$:
$\chi^2 = 237$ ($\chi^2/{\rm dof}=1.05$): for $\Omega_3 = -3.3\pm0.6$,
$\Omega_4 = 1.11\pm0.25$, yielding $\Omega_2=3.2\pm0.7$.

\end{itemize}

Just as this analysis was being completed, the three-year WMAP data (WMAP3)
were released \cite{wmap3}. Assuming a flat Universe, the WMAP3 data by
themselves yield within the $\Lambda$CDM model the result $\Omega_{\rm
M}=0.238^{+0.030}_{-0.041}$. This central value is slightly lower than our
result, but the error is larger, and the two results are compatible within
their errors. The WMAP collaboration has combined their data with other CMB
data, and the most precise result, obtained by combining WMAP3 with the 2dF
Galaxy Redshift Survey, is $\Omega_{\rm M}=0.236^{+0.016}_{-0.024}$, which
differs from our combined result by about $2\sigma$. On the other hand,
combining WMAP3 with large-scale structure data gives larger central values of
$\Omega_{\rm M}$. The smallest errors are those obtained by combining WMAP3
with the Sloan Digital Sky Survey LRG data set:  $\Omega_{\rm
M}=0.267^{+0.018}_{-0.025}$ and with the CFHTLS lensing data: $\Omega_{\rm
M}=0.299^{+0.019}_{-0.025}$. These results straddle our combined fit to the
supernova data.

The data on baryon acoustic oscillations (BAO) provide complementary
information \cite{linder, baryon}. Within the $\Lambda$CDM model and assuming a
flat Universe, the BAO yield an independent estimate~\cite{baryon} $\Omega_{\rm
M}=0.273\pm0.024$, which is completely compatible with the value we obtained
from the supernova data. Combining the two results, we find $\Omega_{\rm
M}=0.274\pm0.014$. We leave for a future occasion the exploration of the
constraints imposed by the BAO data on the Q-cosmology models. When confronting
the BAO data with the Q-cosmology model, it will be desirable also to formulate
more precisely the predictions of this model, which is also an important
subject for future work. There is one important issue which should be stressed
at this point. The exotic matter scaling: $a^{-\delta}$ with $\delta \sim 4$ of
the Q-cosmology model at late eras, if valid at earlier times, would have
dominated the epochs characterised by redshifts $z > 10$. As a result, the
positions of the peaks of the Baryon Acoustic Oscillations would have been
modified, leading to stringent constraints on the models. However, in such a
scenario the phenomenological observables currently used in analyses of the
BAO~\cite{baryon} would be inapplicable to Q-cosmology models. We hope to come
back to this issue in a future publication.

\begin{ack}

N.E.M.\ wishes to thank G.~Diamandis, V.~Georgalas and A.~B.~Lahanas for
discussions, and the Physics Department of Athens University for its
hospitality during the last stages of this work. The work of N.E.M.\ is
partially supported by funds made available by the European Social Fund (75\%)
and National (Greek) Resources (25\%) - EPEAEK~B - PYTHAGORAS. V.A.M.\
acknowledges support by the European Union through the RTN contract:
HPRN-CT-2002-00292, \emph{The Third Generation as a Probe for New Physics}. The
work of D.V.N.\ is supported by D.O.E.\ grant DE-FG03-95-ER-40917.

\end{ack}



\end{document}